\documentstyle[emulateapj]{article}

\def\Ha{H$\alpha$}
\newcommand\hii{\ion{H}{2}}
\newcommand\hi{\ion{H}{1}}
\newcommand\etal{et\thinspace al.~}

\newcommand\msol{\rm\,M_\odot}

\newcommand\ergs{{\rm\,erg\,s^{-1}}}
\newcommand\kms{{\rm\,km\,s^{-1}}}
\def\cc{{\rm\,cm^{-3}}}
\def\ccK{{\rm\,cm^{-3}\ K}}

\def\spose#1{\hbox to 0pt{#1\hss}}
\def\simpropto{\mathrel{\spose{\raise 3pt\hbox{$\propto$}}
     \lower 3.0pt\hbox{$\sim$}}}

\received{26 December 2002}
\accepted{}

\slugcomment{Accepted 31 March 2004 to the Astrophysical Journal}

\lefthead{M. S. Oey \& G. Garc\'\i a-Segura}
\righthead{Ambient Pressure and Superbubble Evolution}

\begin{document}

\title{
Ambient interstellar pressure and superbubble evolution
}
\author{M. S. Oey}
\affil{Department of Astronomy, 830 Dennison Building, 
	University of Michigan, Ann Arbor, MI\ \ \ 48109-1090}
\author{\smallskip and \\ G. Garc\'\i a-Segura}
\affil{Instituto de Astronom\'\i a, Universidad Nacional Autonoma de
	M\'exico, Apartado Postal 877, Ensenada, 22830 Baja
	California, Mexico}

\begin{abstract}
High ambient interstellar pressure is suggested as a possible
factor to explain the ubiquitous observed growth-rate
discrepancy for supernova-driven superbubbles and stellar wind bubbles.
Pressures of $P/k\sim 10^5\ \ccK$ are plausible for regions with high
star formation rates, and these values are intermediate between the
estimated Galactic mid-plane pressure and those observed in starburst
galaxies.  High-pressure components also are commonly seen in Galactic
ISM localizations.  We demonstrate the sensitivity of shell
growth to the ambient pressure, and suggest that superbubbles
ultimately might serve as ISM barometers.
\end{abstract}


\section{Introduction}

Mechanical feedback from supernovae (SNe) and stellar winds is an
important driver of galactic evolutionary processes.  It affects the
phase balance and physical conditions of the interstellar medium
(ISM), which in turn determine star-forming conditions, galactic
chemical evolution, and properties of the intergalactic medium.
The standard paradigm for mechanical feedback is based on the model for
adiabatic evolution of the shells and superbubbles (e.g., Pikel'ner
1968; Weaver {\etal}1977; Mac Low \& McCray 1988) that are the direct
consequence of SN and stellar wind action. 

While this model for pressure-driven superbubbles is broadly
consistent with observations spanning scales from individual stellar
wind bubbles to galactic superwinds, nagging problems
persist in comparisons with observations (e.g., see Oey 2004 for a
review).  Specifics regarding the energy budgets, fate of the
shock-heated $10^6$ K gas, and later-stage evolution are lacking and
have profound consequences for galactic evolution.

One problem that is empirically well-established is the result that most
superbubbles apparently grow more slowly than expected.  This has been
observed in individual stellar wind bubbles such as Wolf-Rayet nebulae
(Treffers \& Chu 1982; Garc\'\i a-Segura \& Mac Low 1995; Drissen
{\etal}1995), as well as in superbubbles powered by OB associations
(e.g., Oey 1996$a$; Oey \& Smedley 1998; Brown {\etal}1995; Saken
{\etal}1992).  This growth-rate discrepancy has been identified in
young, nebular shell systems, in which the parent OB association is
still present; thus the input mechanical power is well-constrained.
The discrepancy is seen both in objects that show no evidence of
previous supernova activity, and in ones where one or two supernovae
have already exploded (Oey 1996$a$; hereafter O96).

For constant  input mechanical power $L$ and ambient number density
$n$, the evolution of the shell radius is given by (e.g., Castor,
McCray, \& Weaver 1975),
\begin{equation}\label{eqR}
R = 68.9\ (L_{38}/n)^{1/5}\ t_6^{3/5}\ \ \rm pc \quad ,
\end{equation}
where $L_{38}$ is $L$ in units of $10^{38}\ \ergs$, and $t_6$ is age of the
bubble in Myr.  The shell expansion velocity $v$ is the time
derivative of equation~\ref{eqR}.  One possible solution to 
the growth-rate discrepancy suggests that the input parameter $L/n$
is systematically overestimated.  For eight nebular superbubbles with
well-constrained $R$, $v$, $L$, and $t$, Oey and collaborators (O96;
Oey \& Massey 1995; Oey \& Smedley 1998) showed that $L/n$ would
need to be reduced by a factor of several, perhaps up to an order of
magnitude, to reconcile the observations with prediction.  Since
stellar wind power $L$ is sensitive to the stellar mass, a substantial
uncertainty in $L$ is not unreasonable.  As shown by multi-wavelength
observations of three of the superbubbles (Oey {\etal}2002), the
multi-phase ambient ISM also renders $n$ similarly uncertain.
However, the implication of a {\it systematic} growth-rate discrepancy
remains difficult to explain.

Another favorite candidate to solve the problem is cooling of the hot
interior whose pressure drives the shell growth.  If this scenario is
correct, it implies a departure from the adiabatic evolution.  The
mass within the hot bubble interior is dominated by material
evaporated from the cool shell walls, and could be supplemented by
additional material evaporated and ablated from small clouds that are
overrun by the expanding shocks (Cowie \& McKee 1977; McKee
{\etal}1984; Arthur \& Henney 1996).  This enhanced interior density
would facilitate radiative cooling.  Silich {\etal}(2001) and Silich
\& Oey (2002) also show that the enhanced metallicity caused by SN
explosions and stellar products can further facilitate the cooling,
especially for low-metallicity objects.  However, increased X-ray
luminosities that are expected from enhanced cooling thus far have not
been observed (Chu {\etal}2003; Chu {\etal}1995).

The superbubbles studied by Oey and collaborators are all located in
the Large Magellanic Cloud (LMC).  For these objects, Silich \& Franco
(1999) suggested that the ambient environment and viewing geometry
conspire to yield misleading observed shell dynamics.  They suggest
that the superbubbles are more extended perpendicular to the galaxy's
plane, as would be expected in the plane-stratified density
distribution of disk galaxies (but see also Maciejewski \& Cox 1999).
The elongation of the shells would not 
be apparent because of the LMC's almost face-on orientation.  While
this is an attractive suggestion for the LMC objects, it does not
explain the growth-rate discrepancy seen in Galactic (e.g., Brown
{\etal}1995; Saken {\etal}1992) and M33 (Hunter {\etal}1995) objects.

Nevertheless, it is apparent that the ambient environment plays a
crucial role in the superbubble growth and evolution.  In addition to
the work of Silich \& Franco (1999), other studies have shown that the
shell dynamics are sensitive to the ambient density structure (e.g., Oey \&
Smedley 1998; Mac Low {\etal}1998).  Multi-wavelength observations
also show that the ambient multiphase gas distribution is difficult to
constrain without direct such observations (Oey {\etal}2002).

\section{The interstellar pressure}

\begin{deluxetable}{lcccclllcc}
\tablewidth{0pt}
\footnotesize
\tablecaption{LMC superbubble parameters\tablenotemark{a}\label{tparams}}
\tablecolumns{10}
\noindent
\tablehead{
\colhead{DEM} & \colhead{$R$} & \colhead{$v$} & 
\colhead{$\log Q^0$} &
\colhead{$120\msol$} & \colhead{$85\msol$} & \colhead{$60\msol$} & 
\colhead{$40\msol$} & \colhead{$25\msol$} & \colhead{$20\msol$} \nl
& \colhead{(pc) \tablenotemark{b}} & \colhead{($\kms$) \tablenotemark{c}} &
\colhead{(log s$^{-1}$) \tablenotemark{d}} & 
\colhead{$\tau=3.12$} & \colhead{$\tau =$3.48} & \colhead{$\tau = $4.12} &
\colhead{$\tau=$5.26} & \colhead{$\tau=$7.84} & \colhead{$\tau=9.96$}
}
\startdata
\cutinhead{Pre-SN superbubbles}
L31  & 50 & 30:           & 50.161 & 1   & 0   & 0    & 1 & 4  & 2 \nl
L106 & 30 & $\lesssim 10$ & 49.745 & 0   & 1   & 0    & 2 & 0  & 4 \nl
L226 & 28 & $\lesssim 5$  & 49.403 & 0   & 0   & 1    & 1 & 0  & 1 \nl
\cutinhead{Post-SN superbubbles\tablenotemark{e}}
L25  & 43 & 60:           & 48.459 & 0   & 0   & 0 (1)&0 (1)& 2 & 2 \nl
L50  & 50 & 25            & 49.342 & 0   &0 (1)& 0 (1)& 3  & 1  & 7 \nl
L301 & 53 & 40:           & 50.310 & 0   &0    & 0 (1)& 3  & 3  & 1 \nl
\tablenotetext{a}{Data compiled by O96).  Columns 6--11
represent numbers of stars in each mass bin; expected lifetime in Myr
is shown in the column heading.}  
\tablenotetext{b}{Uncertainty $\sim 10-15$\%.}
\tablenotetext{c}{Objects with ``:'' uncertain to 50\%, but see text;
	others $\sim 20$\%.  See O96 for source references for $v$.}
\tablenotetext{d}{Uncertainty of order a factor of 2.}
\tablenotetext{e}{Values in parentheses show original number of
stars implied by the IMF, from O96.}
\enddata
\end{deluxetable}

Continuing to focus on the ambient environment, this present work now 
investigates the effect of the ambient pressure.  Since most superbubble
growth is eventually expected to become confined by the ambient
interstellar pressure (Oey \& Clarke 1997), this parameter is also
worth examining more closely.  We note that Dopita et al. (1981)
suggested an active pressure mechanism to confine the shell of one LMC
object by invoking exterior ram pressure caused by contraction of 
a surrounding interstellar cloud.  Here, we suggest that the typical
interstellar pressures in some systems may be higher than assumed.

In our earlier work, we used a simple, semi-analytic, 1-D model that
integrates the shell's equations of motion (O96; Oey \& Massey 1995).
For our sample of eight young, nebular
superbubbles, we tailored the model input parameters ($L,\ n,\ t$)
according to the individual, empirically-derived values.  As mentioned
above, these highly constrained models confirm the ubiquitous
growth-rate discrepancy between the observed 
and predicted shell radius and expansion velocity (e.g., O96).  
The model predicts that the interior pressure evolves as (e.g., Weaver
et al. 1977),
\begin{equation}\label{eqPi}
P_i/k = 1.83\times 10^5\ L_{38}^{2/5} n^{3/5} t_6^{-4/5}\ \ccK \quad.
\end{equation}
In our above studies, the ambient
interstellar pressure $P_e$ was estimated as $P_e = \rho c_s^2 /
\gamma.$  The soundspeed $c_s$ was usually estimated as $10\ \kms$ for
ionized nebular gas, provided that $R$ remains smaller than the
Str\"omgren radius; $\rho$ is the mass density; and $\gamma$ is the 
equation of state index, which was taken to be 5/3 for the adiabatic
condition.  For $n$ ranging between 1 and 10 $\cc$, as estimated for
our objects, this yields $P_e/k = 9\times 10^3$ to $9\times 10^4\
\ccK$.

For the Milky Way, the total mid-plane ISM pressure is generally
estimated to be around $P/k \sim 3\times 10^4\ \ccK$.  This includes
roughly equal empirical contributions from the diffuse thermal pressure,
magnetic field pressure, non-thermal velocity field or turbulent
pressure, and cosmic ray pressure; a good discussion is presented by
Slavin \& Cox (1993).  This value is consistent with the constraint
derived by Boulares \& Cox (1990) that the pressure required to
support the weight of the Galactic ISM is $P/k \sim 2.8\times 10^4\ \ccK$.   

In recent years, turbulent velocity structure is becoming more quantified
as one of the major, and perhaps dominant, kinematic properties of
the ISM in star-forming galaxies.  This is linked to a new paradigm
shift for the ISM to less distinct thermal phases, in which cool
clouds are transient, unconfined features, rather than distinct and
well-defined entities (V\'azquez-Semadeni 2002; Kritsuk \& Norman 2002).
This turbulence-dominated view of the ISM implies a strong presence of
components and localizations that are not in pressure equilibrium (Mac
Low {\etal}2001; Kim {\etal}2001).  Such components are now being
ubiquitously identified in thermal pressure
distributions determined for lines of sight to Galactic stars (Jenkins
\& Tripp 2001; Wallerstein {\etal}1995), where these are found to have
thermal $P_{\rm th}/k \gtrsim 10^5\ \ccK$.  Also note that the turbulent
pressure $P_{\rm tb}/k \simeq \rho\sigma_{\rm v}^2/k \sim 7.5\times 10^4\
\ccK$ for the hot, ionized medium (HIM) if we consider values of
$n\sim 0.05\ \cc$ and turbulent velocity dispersion $\sigma_{\rm
v}\sim c_s \sim 100\ \kms$.  The magnetic field pressure also shows
occasional hints for high-value components (e.g., Edgar \& Cox 1993).
On the other hand, the cosmic ray pressure contribution may be less
relevant to confining superbubble growth since the shells may be
transparent to cosmic rays (Slavin \& Cox 1993). 

The above factors apply to the Milky Way, which is a giant disk galaxy
whose midplane ISM pressure should be substantially higher than in a
Magellanic irregular galaxy like the LMC, where all
of our sample objects are located.  Thus, one could argue that $P_e$
estimated by our earlier models are already on the high side of what
might be expected.  On the other hand, several of the pressure terms,
namely, the thermal, turbulent, and cosmic ray pressures, should scale
with star formation rate (SFR) per unit volume.  Thermal pressures
within star forming regions themselves are of order $P_{\rm th}/k \sim
10^5 - 10^6\ \ccK$ (e.g., Malhotra {\etal}2001), and superbubbles 
generally originate within such regions, although presumably they
outgrow and outlive them.  For a high filling
factor or interstellar porosity generated by the superbubble activity, it
is also quite likely that the magnetic pressure is also determined
by the SFR.  The magnetic field strength is correlated with gas
density, yielding localized values of order a few hundred $\mu$G in
Galactic molecular clouds and star-forming regions (Crutcher 1999).
The expansion of superbubbles themselves also compresses the
magnetic field, causing a self-induced impedance (Slavin \& Cox 1992).
On large scales, radio synchrotron measurements for other galaxies are
showing magnetic field strengths of $\sim 15\ \mu$G in actively
star-forming galaxies, and up to $40\ \mu$G in spiral arms
(Beck 2004).  Such values imply magnetic pressures alone of order
$P_B/k \sim 10^5\ \ccK$.  

The LMC has one of the highest SFR per unit volume in the Local Group.
Its interstellar porosity $Q$ is exceeded only by that of IC~10 
(Oey {\etal}2001), although $Q$ for the Milky Way is difficult to
determine.  Taking the SFR consistently estimated from the \hii\
region luminosity function for both the Galaxy (Oey \& Clarke 1997)
and the LMC (Oey {\etal}2001), we obtain porosities of $Q\sim 0.3$ and
1, respectively.  The formal $Q$ for IC~10 is an order of magnitude
higher yet.  Thus it is not unreasonable that the LMC may have total
interstellar pressures that are similar to, or in excess of, those of
the Milky Way.  We can also compare with the central few kpc of
nuclear starburst galaxies like M82, which have empirically determined
$P/k \sim 10^6\ \ccK$ (e.g., Lord {\etal}1996).  If the Milky Way
ISM, with $P/k \sim 10^4\ \ccK$, is typical for ordinary star-forming
galaxies, then galaxies with SFR intermediate between ``ordinary'' and
starburst, like the LMC, should be expected to likewise have
interstellar pressures that are intermediate, namely, of order
$P/k\sim 10^5\ \ccK$.  Equation~\ref{eqPi} shows that such values
would be significant in counteracting the pressure-driven shell growth.

\section{Models of LMC superbubble evolution}

In light of the above, it is worth further exploring the effects of the
ambient pressure parameter on the superbubble evolution.  We return to
the same LMC objects studied by O96, now modeling these with 1-D
hydrodynamical computations using the magneto-hydrodynamic fluid
solver ZEUS-3D (version 3.4).  This code is the updated, 3-D version
of the two-dimensional code ZEUS-2D (Stone \& Norman 1992), an
Eulerian explicit code which integrates the equations of hydrodynamics
for a magnetized ideal gas.  
The code also works efficiently in one dimension, and we perform the
simulations in spherical coordinates, with 1000 zones in the radial
direction $r$.  Taking the time-dependent stellar wind mechanical luminosity 
$L(t) = \frac{1}{2}\, \dot{M}(t)\ v_{\infty}^2$ for each object from
O96, we assume a constant terminal velocity $v_{\infty}
= 3000\ \kms$, yielding a wind density $\rho(t)=\dot{M}(t) /4\pi r^2
v_{\infty}$, where $\dot{M}(t)$ is the wind mass-loss rate.
O96 estimated $L(t)$ from the observed and inferred stellar population
(Table~\ref{tparams}, below) based on empirical relations between wind
properties and spectral type, and evolutionary models.
The time-dependent wind conditions are set within
the first few radial computational zones, centered at the origin.

ZEUS-3D does not include radiation transfer, but we have
implemented a simple approximation to derive the location of the ionization 
front for arbitrary density distributions (see Bodenheimer et al. 1979). 
This is done by assuming that ionization equilibrium 
holds at all times, and that the gas is fully ionized inside the HII region. 
The position of the ionization front  is given by 
$\int n^2(r) r^2 dr \approx Q^0 / 4 \pi \alpha_B$, where $Q^0$ is the
stellar H-ionizing emission rate and $\alpha_B$ is the Case B
recombination coefficient.
 
The models include the Raymond \& Smith (1977) cooling curve above $10^4$ K. 
For temperatures below $10^4$ K, 
the shocked gas region is allowed to cool down with the radiative cooling
curves given by Dalgarno \& McCray (1972) and MacDonald \& Bailey (1981).
Finally, the photoionized gas is always kept at $10^4$ K , so no cooling curve
is applied to the HII regions (unless there is a shock inside the
photoionized region).

We assume an initially homogeneous ISM, which supplies the external pressure.  
Given the existence of large non-thermal ``turbulent'' velocities,
of several km s$^{-1}$, cosmic rays and magnetic fields, 
the $total$ pressure should be obviously a combination of all of them,
as discussed above.  However, to simplify computation, we model the
total ambient pressure as originating entirely from thermal pressure.
Thus, in order to have thermal 
pressures of $P/k = 1 \times 10^5\ \rm cm^{-3}$ K, we adopt for the
ISM density $n \sim 16.7\ \rm cm^{-3}$ and 6000~K for the
temperature.  This gives us a sound speed of $c_{s,\rm isoth} \sim  7~\kms$.
For a second set of models with $P/k= 1 \times 10^4\ \rm cm^{-3}$ K, we
use an ISM density of $n \sim 1.67\ \rm cm^{-3} $ and 6000~K for the
temperature.

\begin{figure*}
\epsscale{2.0}
\plotone{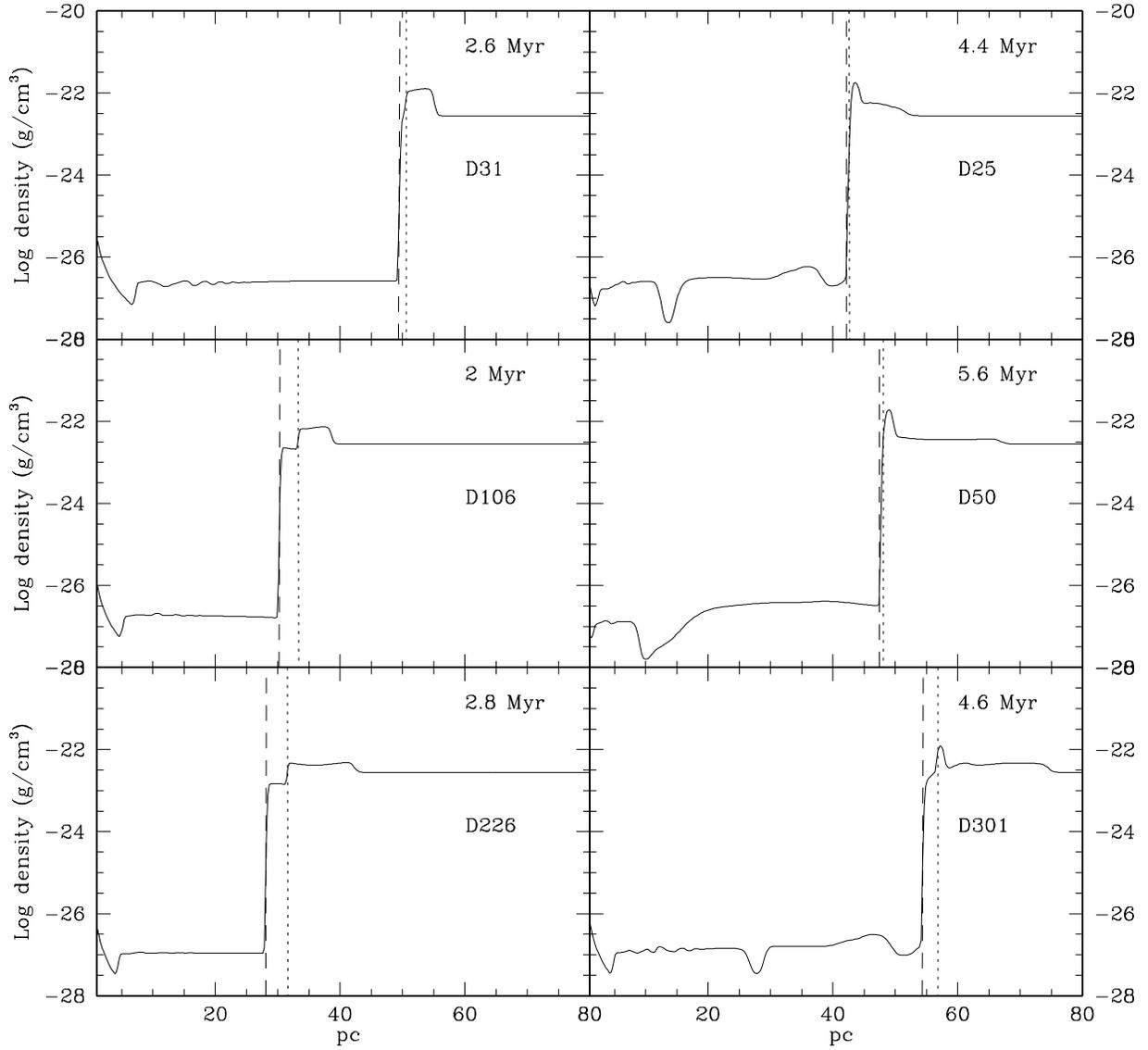}
\vspace*{-2.0 truein}
\caption{Modeled density profiles for the six LMC superbubbles,
assuming an ambient $P/k=1\times 10^5\ \ccK$.  The input stellar populations
are given in Table~1, along with observed parameters.  The observable
nebular shell is delineated by the vertical dashed and dotted line.
\label{fhiP_dens} }
\end{figure*}

\begin{figure*}
\epsscale{2.0}
\plotone{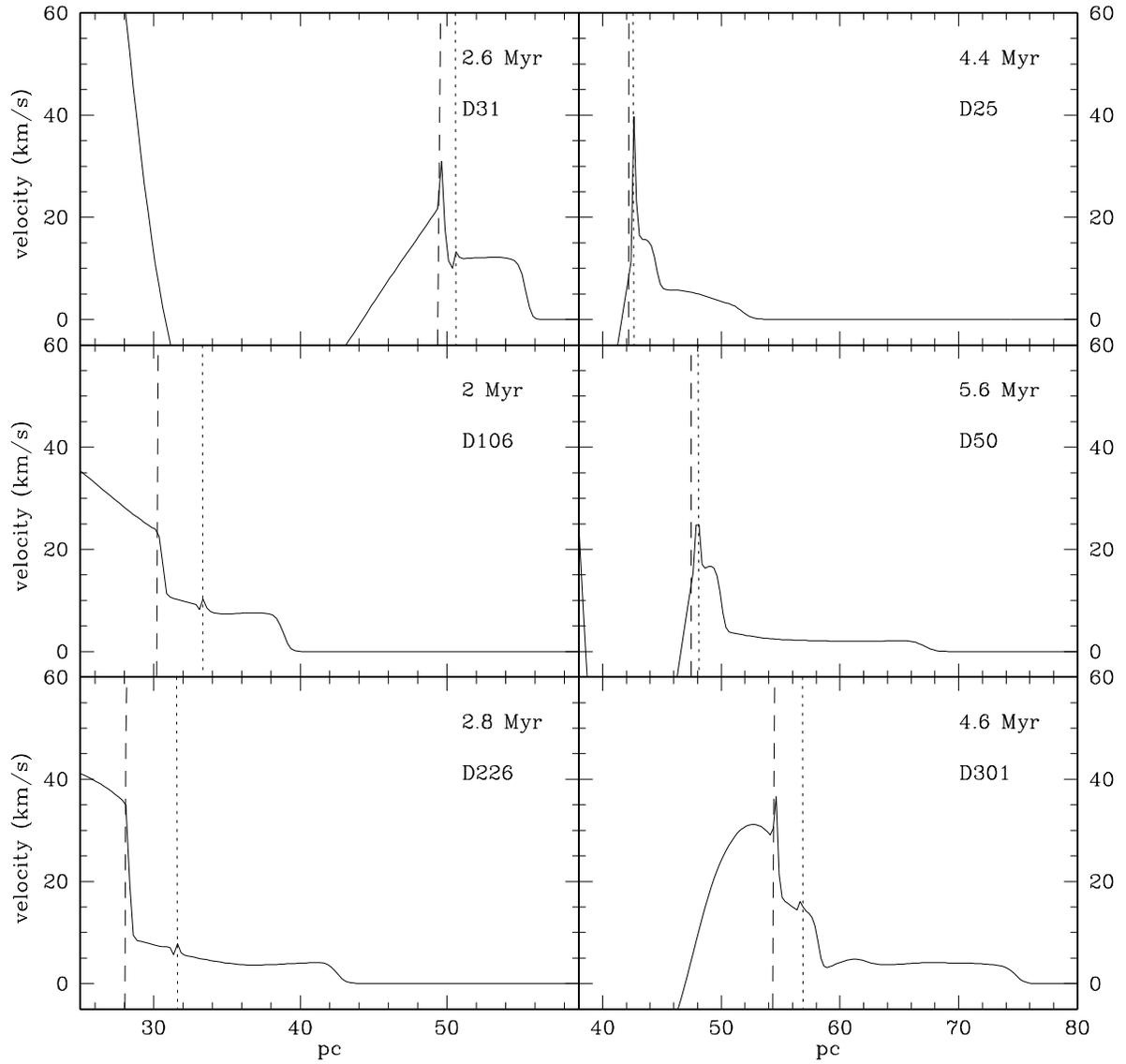}
\vspace*{-2.0 truein}
\caption{Predicted velocity profiles for the models shown in
Figure~\ref{fhiP_dens}, with lines as in Figure~\ref{fhiP_dens}.
\label{fhiP_vel} }
\end{figure*}

Table~\ref{tparams} gives the observed nebular radius $R$, nebular
expansion velocity $v$, and $Q^0$ in columns 2 -- 4, respectively, for
the sample objects.  Columns 5 -- 10 give the observed and 
pre-SN inferred massive star population down to $20\msol$.  O96
describes how the input mechanical power $L(t)$ due to stellar winds and
SNe is estimated from the individual massive star populations as a
function of time.  Figures~1 and 2 of O96 show that $L(t)$ for these
objects typically have early values of $10^{37} - 10^{39}\ \ergs$.
All parameters are listed in Table~\ref{tparams} as measured and
compiled from the literature by O96, except for $Q^0$, which is
estimated by Oey \& Kennicutt (1997).  The objects are divide into
``Pre-SN'' and ``Post-SN'' categories:  the Pre-SN objects show no
enhanced X-ray emission or enhanced [\ion{S}{2}]/\Ha\ ratios, thus are
presumed to have no prior SN activity.  Objects listed as Post-SN
show both enhanced X-ray emission (Chu \& Mac Low 1990; Wang \&
Helfand 1991) and [\ion{S}{2}]/\Ha, indicating recent SN impacts on
the shell walls; the masses of the SN progenitors are estimated from
the stellar initial mass functions (IMF) and included in $L(t)$ for
the models, as done by O96. 

Figure~\ref{fhiP_dens} shows our new models for the gas mass density as a
function of radius, using the input $L(t)$ determined by O96 from the
observed stellar population (Table~\ref{tparams}) of the
individual LMC objects.  These models adopt a high ambient pressure
$P/k = 1\times 10^5\ \ccK$.  As discussed earlier, contributions
from turbulent, magnetic, and multi-phase thermal pressure terms would
presumably contribute to such high ambient values.  The models are
those corresponding to the observed radii in Table~\ref{tparams}, with
Pre-SN objects shown in the left column, and Post-SN objects shown
on the right.  Figure~\ref{fhiP_vel} shows the gas velocity as a
function of radius for the same models.  The location of the observable
ionized nebula is shown between the vertical dashed and dotted line in
both Figures.

Comparison with Table~\ref{tparams} shows that the high-pressure
models are in good agreement with the observed parameters in all cases:
the predicted ionized radius, expansion velocity, and age are all
reasonably consistent with the observed values.  In Figure~\ref{fHa} we
reproduce \Ha\ images of the superbubbles from Figure~1 of Oey
(1996$b$).  Figure~\ref{fHa} also shows that the putative Pre-SN
objects DEM L106 and L226 show thicker nebular shells, of order 10\%
of $R$, whereas 
the post-SN objects show compressed, filamentary morphology.  DEM L31,
presumed to be Pre-SN, is also quite filamentary, and the model 
also reproduces this structure, owing primarily to its high input
power (Table~\ref{tparams}).  This is fully consistent with the
predicted morphology of the ionized regions in the models.  

\begin{figure*}
\epsscale{2.0}
\plotone{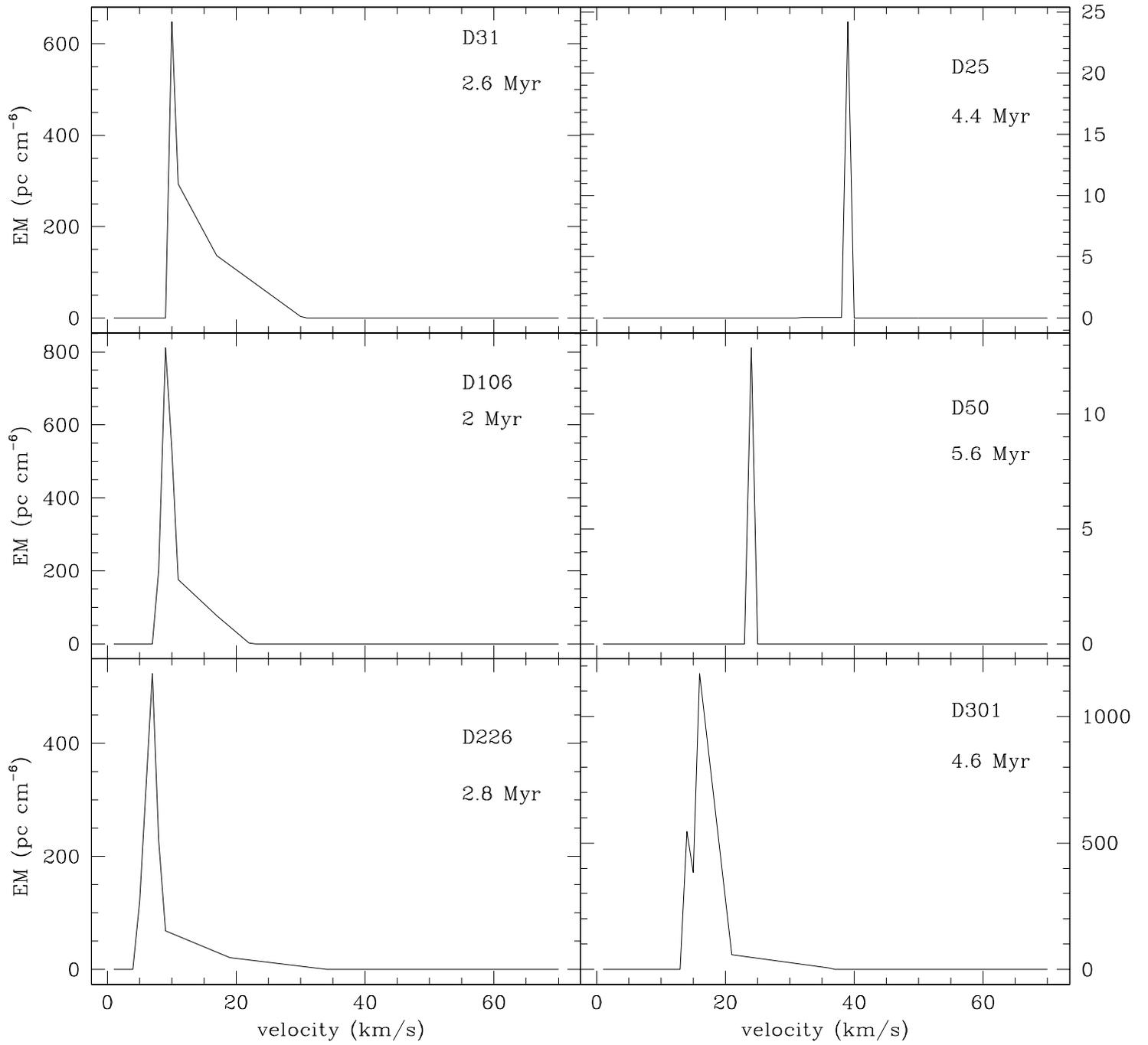}
\vspace*{-2.0 truein}
\caption{Predicted positive \Ha\ line profiles for the models shown in
Figure~\ref{fhiP_dens}.  Note that the total line profile will be
symmetric around the origin (systemic velocity).
\label{fhiP_velprof} }
\end{figure*}

The velocity profiles shown in Figure~\ref{fhiP_vel} also imply
observed velocities that are reasonably consistent with the data
(Table~\ref{tparams}).  The latter are especially complex for the
most filamentary objects, showing velocity components that vary by
factors of 2--3.  This is consistent with the extreme gradients in 
both density and velocity seen in the models.
Figure~\ref{fhiP_velprof} shows the resulting
predicted, positive velocity profiles with respect to the systemic 
velocities, integrated through the centers of the superbubbles. 
DEM L31 and DEM L301 show a significant range in velocity, and
maximum values consistent with those observed.  Predicted velocities
for the remaining objects are also consistent with the observed upper
limits.  We note that both observations and models are likely to be
sensitive to details of geometry and ionization in the post-SN objects,
since complex shock structures are generated on short timescales.
Thus it is unsurprising that the model velocity profiles tend to be
more simplistic than the observations.  We also see that the models for 
DEM L25 and DEM L50 imply that the ionization front does not penetrate
the high-density shell (Figure~\ref{fhiP_dens}); however, observations
of these objects do show high-density, filamentary structure
that is fully ionized for DEM L25, and mostly ionized for DEM L50 (Oey
et al. 2002).  Oey \& Kennicutt (1997) find excess in the nebular
emission, by factor of a few, beyond what can be attributed to stellar
photoionization.  Thus, shock excitation is likely to enhance the
ionized mass and radius of these objects.  Nevertheless, the broad
agreement in the dominant shell parameters across the sample demonstrates
that {\it a high ambient pressure alone can solve the growth-rate
discrepancy} that is widely observed in such objects.

\begin{figure*}
\epsscale{1.5}
\plotone{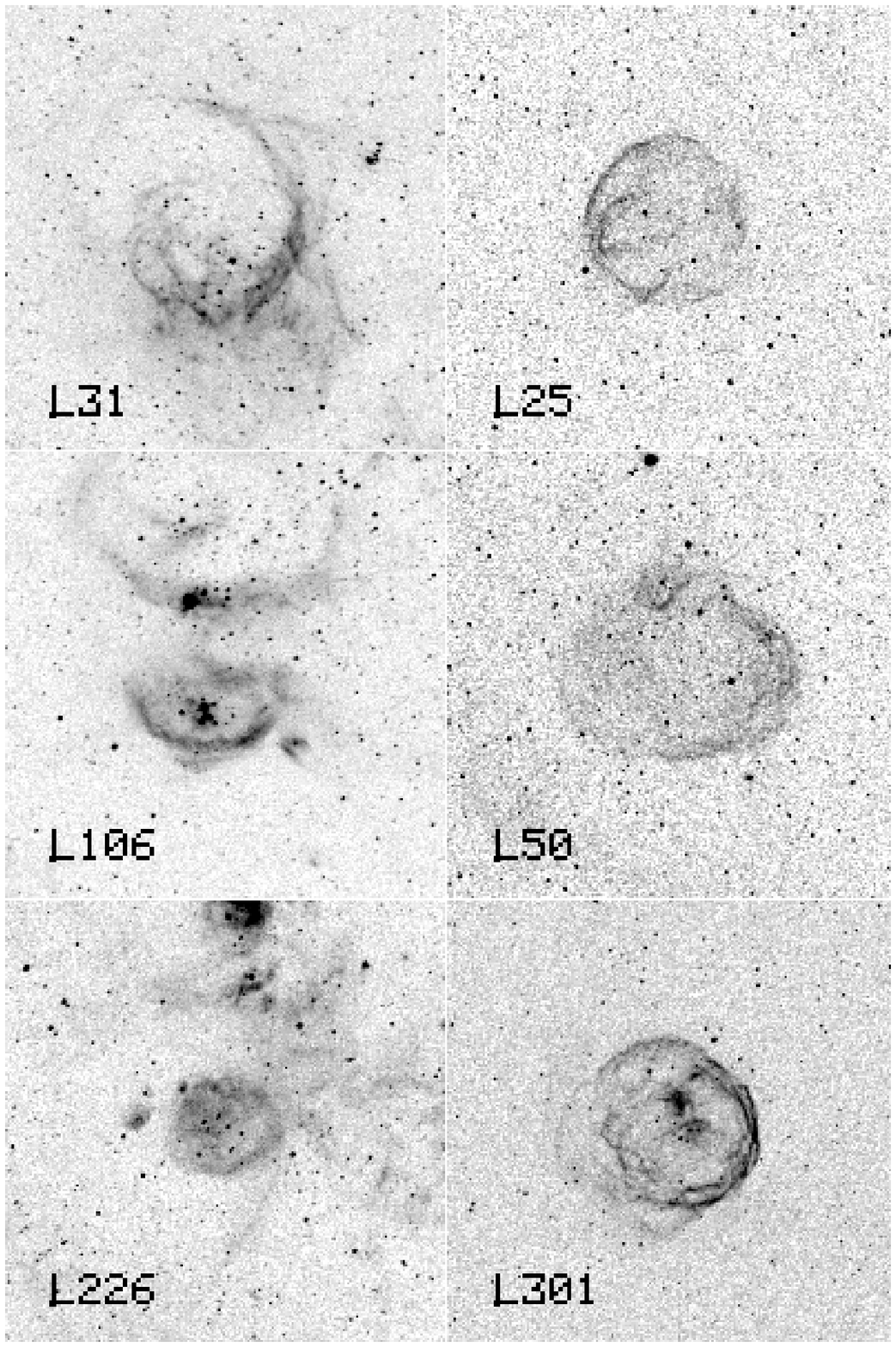}
\caption{\Ha\ images of the LMC superbubbles, from Oey
(1996$b$).  Each image is $16.\arcmin 67$ square.
\label{fHa} }
\end{figure*}

\begin{figure*}
\epsscale{1.0}
\plotone{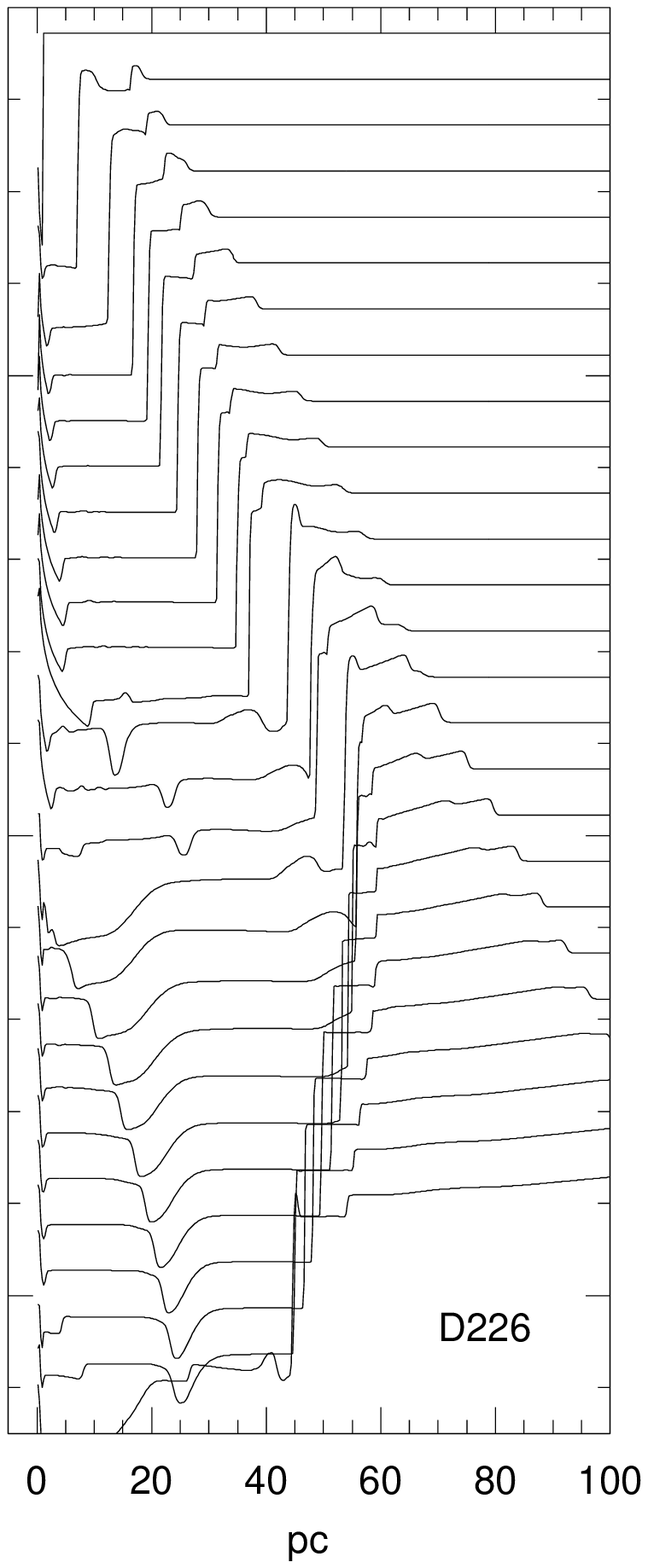}
\plotone{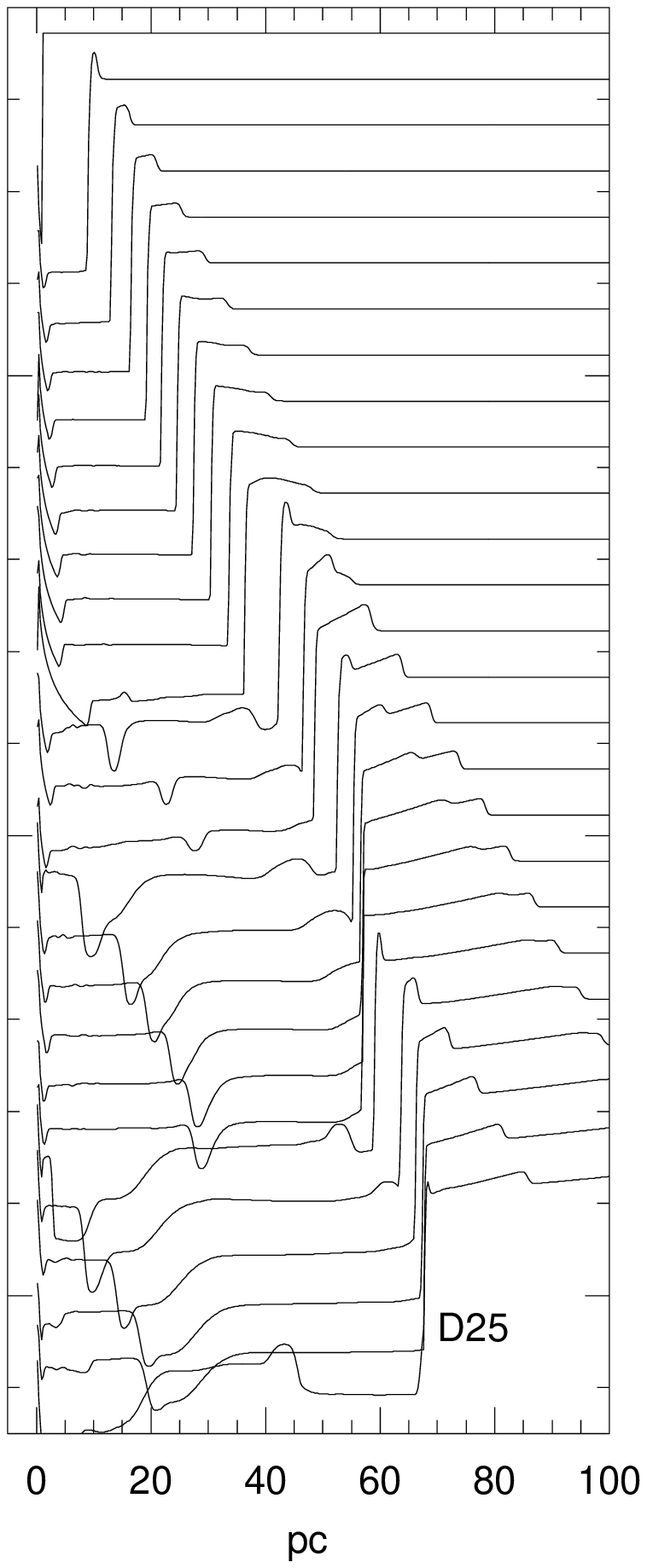}
\vspace{2.0in}
\caption{Predicted evolution of the density profiles for DEM L226 and
DEM L25.  Modeled radial profiles are shown at intervals of $4\times
10^5$ yr, over a total of 10 Myr.  The observed $R$ are consistent
with the models at ages of 2.8 Myr and 4.4 Myr for DEM L226 and DEM
L25, respectively.
\label{fhiP_evol} }
\end{figure*}

Figure~\ref{fhiP_evol} shows the density profile evolution modeled for
the parameters of DEM L226 and DEM L25.  The former is representative
of Pre-SN, low-$L$ objects; and the latter is representative of Post-SN,
higher-$L$ objects.  The curves show a sequence of density profiles
over 10 Myr at $4\times 10^5$ yr intervals, from top to bottom.
We see that the evolution of both objects is similar, but that the
model for DEM L226 does not generate enough internal pressure to
maintain shell growth within the modeled time frame:  the central
cavity is seen to become pressure-confined, and to start collapsing.
On the other hand, the predicted evolution for DEM L25 does maintain
shell growth within the modeled time frame.  In both models, the shell
thickens and diffuses as the expansion falls below the soundspeed,
thereby allowing the shock front and shell to dissipate.  This is
consistent with behavior found by Slavin \& Cox (1992) and  Garc\'\i
a-Segura \& Franco (1996) for similar pressure-confined models.  At
similar, late, subsonic/pressure-confined stages, the structure is
vulnerable to disruption by ambient turbulence, especially if the
latter is an important contributor to a high-pressure situation.  In
the meantime, both models in Figure~\ref{fhiP_evol} also clearly show
compression in the shell caused by supersonic SN impacts around 4.1
Myr.  Additional, subsequent SN impacts also can be seen in the
models.  We do note that DEM L25 is the only object whose age in
Figure~\ref{fhiP_dens} is somewhat inconsistent with the predicted
stellar population.  For a standard IMF, one 40$ \msol$ star is
expected to have existed, which presumably has since exploded,  
implying an age of 5.3 -- 7.8 Myr.  The evolutionary sequence in
Figure~\ref{fhiP_evol} shows $R\sim 53$ pc, vs the observed 43 pc,
at that stage.  This is not a large discrepancy, and the existence of
the putative $40\ \msol$ SN progenitor is also extremely uncertain in
view of the stochastic effects in the IMF.

\begin{figure*}
\epsscale{2.0}
\plotone{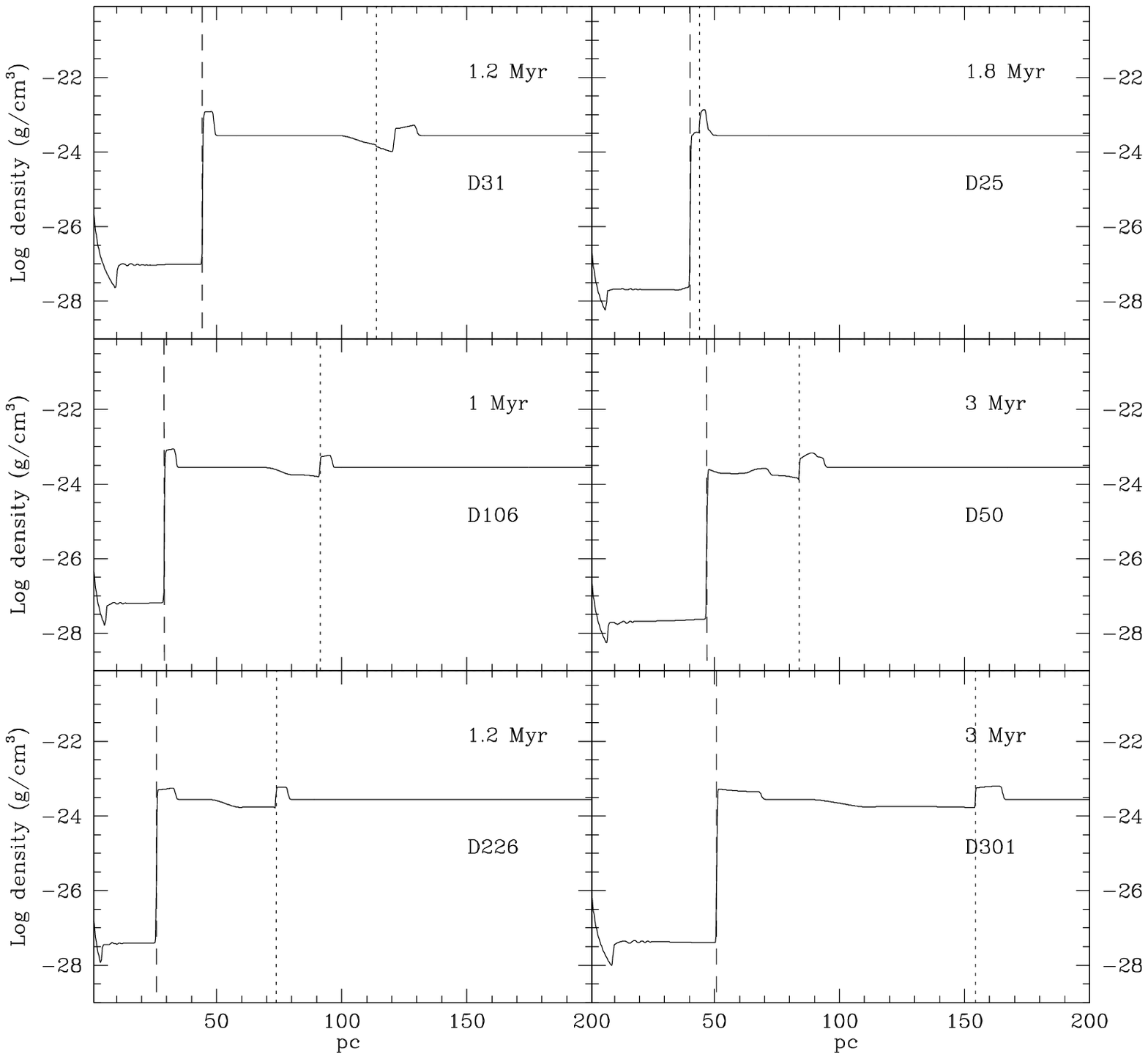}
\vspace*{-2.0 truein}
\caption{Modeled radial density profiles for the same objects,
assuming an ambient $P/k=1\times 10^4\ \ccK$, for $R$ corresponding to
the observed values.  Lines are as in Figure~\ref{fhiP_dens}.
\label{floP_dens} }
\end{figure*}

\begin{figure*}
\epsscale{2.0}
\plotone{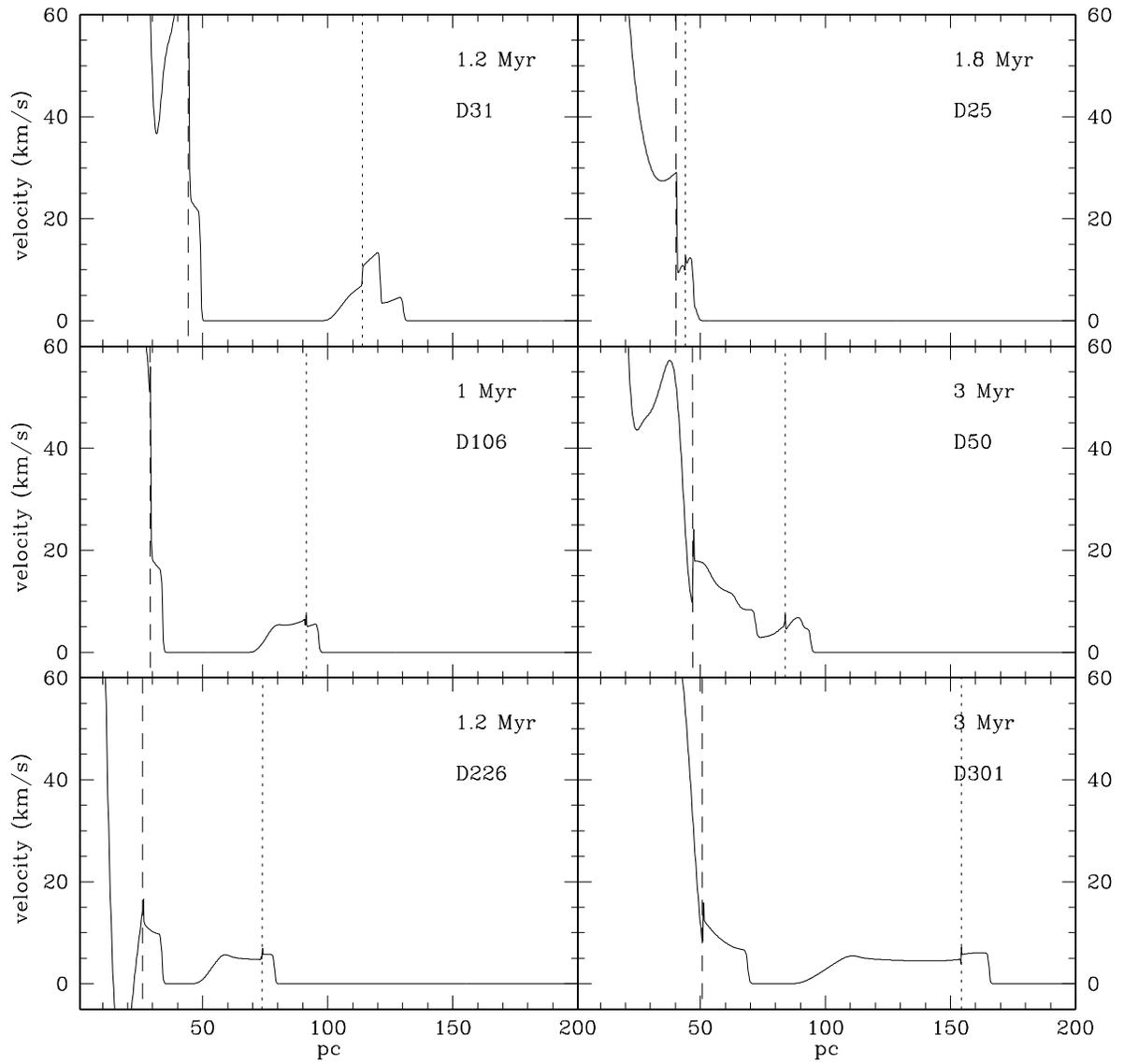}
\vspace*{-2.0 truein}
\caption{Predicted velocity profiles for the models shown in
Figure~\ref{floP_dens}. 
\label{floP_vel} }
\end{figure*}

\begin{figure*}
\epsscale{2.0}
\plotone{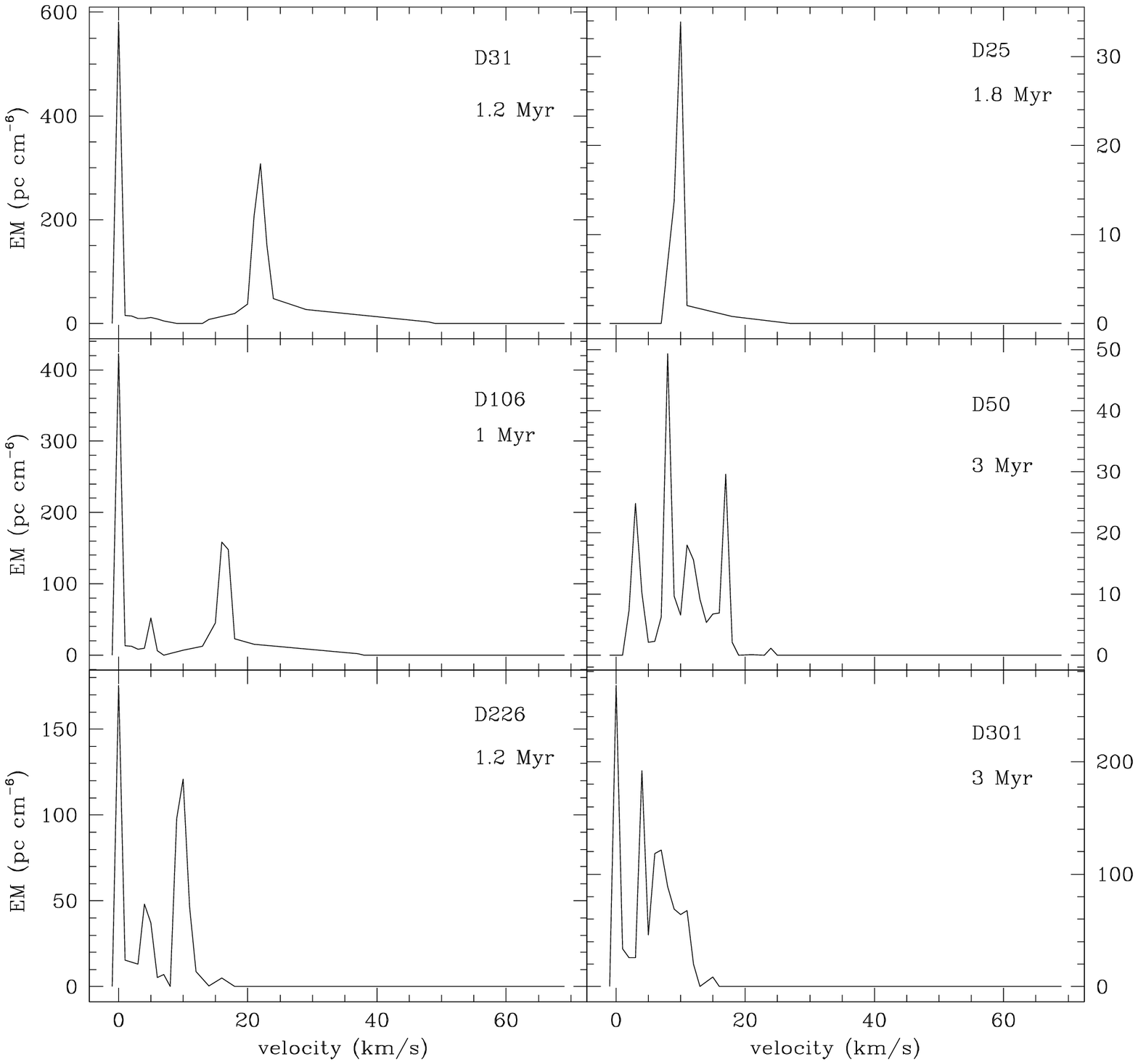}
\vspace*{-2.0 truein}
\caption{Predicted positive \Ha\ line profiles for the models shown in
Figure~\ref{floP_dens}. 
\label{floP_velprof} }
\end{figure*}

We also compare with similar models assuming a lower ambient pressure
of $P/k = 1\times 10^4\ \ccK$, implemented as described above.
Figures~\ref{floP_dens} and \ref{floP_vel} show the 
models corresponding to the observed radii for the same six objects.
While this ambient $P/k$ value is more consistent with standard
expectations, it is apparent that the models are in poor agreement
with the data.  The collective ages are systematically too
young, thus demonstrating the ubiquitous growth-rate discrepancy, and
the predicted shell parameters are more difficult to reconcile with
the observations.  Since none of the objects have reached the SN
stage, the Post-SN objects retain thick, extended shells.  Indeed,
all but one of the objects are predicted to have a surrounding ionized
halo that is at most a factor of 3 lower in density than the inner
shell; these halos are not observed.  Another especially interesting
feature in these models is the appearance of a double shell structure:
a second, neutral shell is seen at roughly 2 -- 3 times the radius of
the ionized inner shell.  This outer shell results from the expansion
of the photoionized \hii\ region in a D-type ionization front.
Oey {\etal}(2002) 
presented \hi\ observations of the three post-SN objects (DEM L25, DEM
L50, and DEM L301), at a spatial resolution of $50\arcsec$ (12.5 pc);
they found no apparent evidence of such secondary shells.  The
existence of neutral secondary shells could potentially offer a
diagnostic of lower ambient pressures.  Figures~\ref{floP_vel} and 
~\ref{floP_velprof} also show the predicted velocity structure and
\Ha\ line profiles for these young objects, which, as expected, are
less consistent with the observations in Table~\ref{tparams}, than
are Figures~\ref{fhiP_vel} and \ref{fhiP_velprof}.  

\section{Conclusion}

Our models clearly show that increasing the ambient 
interstellar pressure by an order of magnitude, from $P_e/k = 1\times
10^4$ to $1\times 10^5\ \ccK$, can impede the shell growth to a degree
that could fully account for the observed growth-rate discrepancy.
In \S 2, we presented arguments that such high interstellar pressures
could exist, especially based on the dependence of $P_e$ on
star-formation rate.  While other factors mentioned in \S 1, namely,
overestimated $L/n$, elevated radiative cooling, and viewing geometry,
could  all be additional factors that contribute to the growth-rate
discrepancy, we note that the multi-phase gas morphology is more
consistent with high interstellar pressure dominating this effect.

Finally, as noted by Oey \& Clarke (1997), the assumed global value of
$P_e$ plays a critical role in determining the characteristic final
sizes of old, SN-dominated superbubbles, and hence, the superbubble
size distribution, which is dominated by pressure-confined shells.
This, in turn, determines the interstellar porosity and filling factor
of the hot, ionized medium in star-forming galaxies.  With adequate
clarification in the superbubble evolution process and input
parameters, the superbubble sizes, kinematics, and morphologies could
potentially provide barometers for the interstellar pressure.  These
diagnostics could be especially useful in other galaxies, which have
fewer available empirical pressure indicators than the Milky Way.

\acknowledgments
We are grateful to the anonymous referee, whose comments led to a much
stronger analysis of this problem.  Thanks also to Dave Strickland for
comments on the manuscript and for prodding us to finally carry
out this work.  We are also pleased to acknowledge comments and
discussions with Don Cox and Robin Shelton.  Much of this work was
carried out by the PI at Lowell Observatory.


\end{document}